# The power of linear programming for valued CSPs: a constructive characterization

Vladimir Kolmogorov
Institute of Science and Technology Austria
vnk@ist.ac.at

**Abstract**

A class of valued constraint satisfaction problems (VCSPs) is characterised by a valued constraint language, a fixed set of cost functions on a finite domain. An instance of the problem is specified by a sum of cost functions from the language with the goal to minimise the sum.

We study which classes of finite-valued languages can be solved exactly by the *basic linear programming relaxation* (BLP). Thapper and Živný showed [20] that if BLP solves the language then the language admits a *binary commutative fractional polymorphism*. We prove that the converse is also true. This leads to a necessary and a sufficient condition which can be checked in polynomial time for a given language. In contrast, the previous necessary and sufficient condition due to [20] involved infinitely many inequalities.

More recently, Thapper and Živný [21] showed (using, in particular, a technique introduced in this paper) that core languages that do not satisfy our condition are NP-hard. Taken together, these results imply that a finite-valued language can either be solved using Linear Programming or is NP-hard.

## 1 Introduction

We consider a particular linear programming relaxation of a class of optimization problems called in this paper *basic LP* (BLP). This relaxation has been studied extensively in various domains, especially for objective functions with unary and binary terms. Researchers analyzed its properties, developed efficient algorithms for (approximately) solving this LP, and applied to large-scale practical problems [18, 13, 22, 10, 23, 5, 15, 19, 1, 11, 17].

It has been long recognized that for some classes of optimization problems (e.g. for submodular functions on a chain) BLP relaxation is guaranteed to be tight (i.e. the integrality gap is zero), and allows to solve the problem exactly. A natural question is whether it is possible to characterize such classes.

One possible way to pose the problem formally is to use the framework of *Valued Constraint Satisfaction Problems (VCSPs)* [4]. In this framework a class of allowed objective functions is specified by a *language* $\Gamma$, which is a collection of cost functions over a fixed domain $D$. We say that BLP *solves* $VCSP(\Gamma)$ if the relaxation is tight for all functions that can be expressed as a sum of functions from $\Gamma$ with overlapping sets of variables.

A major step in the characterization of languages that can be solved by BLP has been recently made by Thapper and Živný [20]. They gave a sufficient condition for BLP to solve $VCSP(\Gamma)$ that covers many known languages such as (1) submodular functions on arbitrary lattices; (2) $k$-submodular functions; (3) weakly (and hence strongly) tree-submodular functions on arbitrary trees.

Thapper and Živný also presented a necessary and sufficient condition for BLP to solve $VCSP(\Gamma)$. However, their characterization has one drawback: it involves infinitely many inequalities, which leaves an open question whether checking the condition is a decidable problem for a given finite language $\Gamma$.

We resolve this question affirmatively for finite-valued languages $\Gamma$. As our **main contribution**, we show that BLP solves such $\Gamma$ iff $\Gamma$ admits a *fractional symmetric polymorphism of arity $k$* for some $k \geq 2$. We prove this result using a mixture of algebraic tools and techniques from Linear Programming.

Very recently, Thapper and Živný [21] showed (using, in particular, a technique introduced in this paper) that a core language that does not satisfy our condition is NP-hard. It follows from [20, 21] and from our results that a finite-valued language can either be solved by a Linear Programming or is NP-hard.

**Related work** Kun et al. [14] studied the BLP relaxation for CSP problems and for robust approximatility of Max-CSPs. They showed, in particular, that BLP robustly decides a CSP language iff it has *width 1*. Width-1 CSPs were introduced in [7]. A simple characterization of such CSPs was given in [6].

Our work heavily exploits the notion of *fractional polymorphisms* [2]. Fractional polymorphisms is a generalization of *multimorphisms* [4]. It is known that they can characterize all tractable VCSPs [2].

We also mention the work of Raghavendra on analyzing SDP relaxations [16]. Under the assumption of the *unique games conjecture* [9], it was shown that the basic SDP relaxation solves all tractable finite-valued VCSPs (without a characterization of the tractable cases). Furthermore, results in Chapters 6 and 7 of [16] imply that the basic SDP relaxation solves languages that admit a *cyclic fractional polymorphism* of some arity $m \geq 2$. If was not clear whether the SDP relaxation can solve exactly more languages compared to the BLP relaxation. Results in [20, 21] and our results imply that this is not the case (assuming that $P \neq NP$).

## 2 Background and statement of the results

Let $D$ be a finite domain. A *finite-valued language* $\Gamma$ is a set of cost functions $f : D^n \to \mathbb{Q}$ where arity $n \geq 1$ may be different for different functions $f \in \Gamma$. The argument of $f$ is called a *labeling*.

**Definition 1.** *An instance $\mathcal{I}$ of the valued constraint satisfaction problem (VCSP) is a function $D^V \to \mathbb{Q}$ given by*
$$Cost_{\mathcal{I}}(x) = \sum_{t \in T} f_t(x_{v(t,1)}, \ldots, x_{v(t,n_t)})$$
*It is specified by a finite set of nodes $V$, finite set of terms $T$, cost functions $f_t : D^{n_t} \to \mathbb{Q}$ of arity $n_t$ and indices $v(t,k) \in V$ for $t \in T, k = 1, \ldots, n_t$. A solution to $\mathcal{I}$ is a labeling $x \in D^V$ with the minimum cost. Instance $\mathcal{I}$ is called a $\Gamma$-instance if all terms $f_t$ belong to $\Gamma$.*

The class of optimization problems consisting of all $\Gamma$-instances is referred to as $VCSP(\Gamma)$. Language $\Gamma$ is called tractable if $VCSP(\Gamma')$ can be solved in polynomial time for **each** finite $\Gamma' \subseteq \Gamma$. It is called NP-hard if $VCSP(\Gamma')$ is NP-hard for **some** finite $\Gamma' \subseteq \Gamma$.

One way to tackle a VCSP instance is to formulate and solve a convex relaxation of the problem. Two examples are *basic LP relaxation* and *basic SDP relaxation*, as they are called in [20] and [16] respectively.

The basic LP relaxation will be of particular relevance to this paper. Following [20], we say that basic LP *solves* $VCSP(\Gamma)$ if for any instance $\mathcal{I}$ from $VCSP(\Gamma)$ the optimal value of the relaxation equals $\min_x Cost_{\mathcal{I}}(x)$.

We will study which languages $\Gamma$ are solved by basic LP. We will do it indirectly by relying on the characterization of [20]. The formulation of basic LP will not be used, and so we omit it in order to avoid unnecessary notation. We refer interested readers to [20] for details.

**Fractional polymorphisms** We denote $\mathcal{O}^m$ to be the set of operations $g : D^m \to D$. A *fractional polymorphism* of arity $m$ is a probability distribution $\omega$ over $\mathcal{O}^m$, i.e. a vector with components $\omega(g) \geq 0$ for $g : D^m \to D$ that sum to 1. Language $\Gamma$ is said to *admit* $\omega$ if every cost function $f \in \Gamma$ of arity $n$ satisfies

$$\sum_{g \in \mathcal{O}^m} \omega(g) f(g(x^1, \ldots, x^m)) \leq f^m(x^1, \ldots, x^m) \qquad \forall x^1, \ldots, x^m \in D^n \tag{1}$$

where function $f^m$ is defined via $f^m(x^1, \ldots, x^m) = \frac{1}{m}(f(x^1) + \ldots + f(x^m))$. We view labelings in $D^n$ as column vectors; given $m$ such columns, operation $g : D^m \to D$ produces a new column as shown on the right.

| $x^1$ | $\ldots$ | $x^m$ | $g(x^1, \ldots, x^m)$ |
|---|---|---|---|
| $x_1^1$ | $\ldots$ | $x_1^m$ | $g(x_1^1, \ldots, x_1^m)$ |
| $\ldots$ | $\ldots$ | $\ldots$ | $\ldots$ |
| $x_n^1$ | $\ldots$ | $x_n^m$ | $g(x_n^1, \ldots, x_n^m)$ |

Operation $g \in \mathcal{O}^m$ is called *symmetric* if it is invariant with respect to any permutation of its arguments: $g(a_1, \ldots, a_m) = g(a_{\pi(1)}, \ldots, a_{\pi(m)})$ for any permutation $\pi : [1, m] \to [1, m]$ and any $(a_1, \ldots, a_m) \in D^m$. It is called *cyclic* if $g(a_1, a_2, \ldots, a_m) = (a_2, \ldots, a_m, a_1)$ for any $(a_1, \ldots, a_m) \in D^m$. Note, in the case $m = 2$ both definitions coincide. A fractional polymorphism $\omega$ is called symmetric (cyclic) if all operations in $supp(\omega)$ are symmetric (cyclic). As usual, $supp(\omega)$ denotes the support of distribution $\omega$: $supp(\omega) = \{g \in \mathcal{O}^m \mid \omega(g) > 0\}$.

**Generalized fractional polymorphisms** Let $\mathcal{O}^{m \to k}$ be the set of mappings $\mathbf{g} : D^m \to D^k$. A mapping $\mathbf{g} \in \mathcal{O}^{m \to k}$ can also be viewed as a sequence of $k$ operations $\mathbf{g} = (g_1, \ldots, g_k)$ with $g_i \in \mathcal{O}^m$. We define a *generalized fractional polymorphism of arity $m \to k$* as a probability distribution $\rho$ over $\mathcal{O}^{m \to k}$. We say that language $\Gamma$ admits $\rho$ if every cost function $f \in \Gamma$ of arity $n$ satisfies

$$\sum_{\mathbf{g} \in \mathcal{O}^{m \to k}} \rho(\mathbf{g}) f^k(\mathbf{g}(x^1, \ldots, x^m)) \leq f^m(x^1, \ldots, x^m) \qquad \forall x^1, \ldots, x^m \in D^n \tag{2}$$

Equivalently, $\rho$ is a generalized fractional polymorphism of $\Gamma$ of arity $m \to k$ iff vector

$$\omega = \sum_{\mathbf{g}=(g_1, \ldots, g_k) \in \mathcal{O}^{m \to k}} \rho(\mathbf{g}) \cdot \frac{1}{k}(\chi_{g_1} + \ldots + \chi_{g_k}) \tag{3}$$

is a fractional polymorphism of $\Gamma$ of arity $m$.

We can identify fractional polymorphisms of arity $m$ with generalized fractional polymorphisms of arity $m \to 1$. For brevity, we will omit the word "generalized".

We always use the following convention: if $\mathbf{g} = (g_1, \ldots, g_k)$ is a mapping in $\mathcal{O}^{m \to k}$ then we extend it to $m$ labelings $x^1, \ldots, x^m \in D^n$ component-wise, i.e. $[\mathbf{g}(x^1, \ldots, x^m)]_v = \mathbf{g}(x_v^1, \ldots, x_v^m)$ for all $v \in [1, n]$. Thus, $\mathbf{g}(x^1, \ldots, x^m)$ is a sequence of $k$ labelings $(g_1(x^1, \ldots, x^m), \ldots g_k(x^1, \ldots, x^m))$ in $D^n$.

**Fractional polymorphisms and SDP/LP relaxations** It has been shown that fractional polymorphisms can be used for characterizing languages that can be solved exactly by certain convex relaxations. For the SDP relaxation, the following is known; it is implied by results in Chapters 6 and 7 of [16].

**Theorem 2** ([16]). *If $\Gamma$ has a cyclic fractional polymorphism of some arity $k \geq 2$ then the basic SDP relaxation solves VCSP($\Gamma$) in polynomial time.*

The more relevant for our paper case of the LP relaxation has been analyzed in [20]:[1][2]

---
[1] Note, direction (i)⇒(ii) of theorem 3 was proved only for a language $\Gamma$ with finite $|\Gamma|$. With a straightforward continuity argument the theorem can be extended to languages with countable $\Gamma$ (by showing first that the set of $m$-ary fractional polymorphisms of any $\Gamma$ is a compact subset of $\mathbb{R}^{\mathcal{O}^m}$).

[2] Thapper and Živný also present some results for infinite-valued languages; we refer to [20] for details.

**Theorem 3** ([20])**.** *For a finite-valued language $\Gamma$, the basic LP relaxation solves $VCSP(\Gamma)$ (in polynomial time) iff $\Gamma$ admits an m-ary symmetric fractional polymorphism for every $m \geq 2$.*

**Theorem 4** ([20], Theorem 4.4)**.** *If $\Gamma$ admits a k-ary fractional polymorphism $\omega$ such that $supp(\omega)$ generates a symmetric m-ary operation then $\Gamma$ admits an m-ary symmetric fractional polymorphism.*

While the theorems give a necessary and sufficient condition for BLP to solve $VCSP(\Gamma)$, it was not clear whether checking this condition for a given language $\Gamma$ is a decidable problem (we would need to consider infinitely many values of $m$).

Our result given below resolves this question affirmatively.

**Theorem 5.** *Suppose that a finite-valued language $\Gamma$ admits a fractional polymorphism of arity $k \geq 2$ whose support contains at least one symmetric operation. Then $\Gamma$ admits a symmetric fractional polymorphism of every arity $m \geq 2$ (and thus BLP solves $VCSP(\Gamma)$).*

A more recent result implies a dichotomy for finite-valued languages: every language is either tractable (via BLP) or is NP-hard.

**Theorem 6** ([21])**.** *If a finite-valued core language $\Gamma$ does not admit a symmetric fractional polymorphism of arity $2$ then it is NP-hard.*[3]

**STP multimorphisms**  As a minor contribution, we formally prove that if a finite-valued language admits an *STP multimorphism* then it also admits a submodularity multimorphism with respect to some total order on $D$. (This has implications for the complexity classification of *conservative* finite-valued languages.) This result is already known, but to our knowledge a formal proof has not been written down yet. We refer to section 4 for further discussion.

## 3  Proof of theorem 5

We will prove the following result.

**Theorem 7.** *Suppose that a finite-valued language $\Gamma$ over a domain $D$ admits a symmetric fractional polymorphism of arity $m - 1 \geq 2$. Then $\Gamma$ admits a symmetric fractional polymorphism of arity $m$.*

This will imply theorem 5. Indeed, suppose that $\Gamma$ admits a fractional polymorphism of arity $k \geq 2$ which contains a symmetric operation. Induction on $m$ yields that $\Gamma$ admits an $m$-ary symmetric fractional polymorphism for every $m \geq k$ (we need to use theorem 4 for the base case and theorem 7 for the induction step). This also implies the claim for all $m \in [2, k-1]$: it is straightforward to show that if $\Gamma$ admits a symmetric fractional polymorphism of arity $pm$ where $p, m \in \mathbb{N}$ then it also admits a symmetric fractional polymorphism of arity $m$.

We thus concentrate on proving theorem 7. From now on we fix a symmetric fractional polymorphism $\omega$ of $\Gamma$ of arity $m - 1$.

Consider a permutation $\pi$ of size $m$ and a symmetric operation $s \in supp(\omega)$ of arity $m - 1$. For such $\pi$ and $s$ we introduce the following definitions.

---
[3] We refer to [21] for the definition of a core. (It differs from the definition in [8], but [21] shows that the two definitions are equivalent.) The coreness assumption is not a severe restriction: if $\Gamma$ is not a core then there is a polynomial-time reduction between $VCSP(\Gamma)$ and $VCSP(\Gamma')$ for some core language $\Gamma'$ on a smaller domain. It is also not difficult to show that Theorem 6 holds for non-core languages as well.

- For a labeling $\alpha = (a_1, \ldots, a_m) \in D^m$, let $\alpha^\pi \in D^m$ and $\alpha^s \in D^m$ be the following labelings:

$$\alpha^\pi = (a_{\pi(1)}, \ldots, a_{\pi(m)}) \tag{4}$$
$$\alpha^s = (s(\alpha_{-1}), \ldots, s(\alpha_{-m})) \tag{5}$$

where $\alpha_{-i} \in D^{m-1}$ is the labeling obtained from $\alpha$ by removing the $i$-th element.

- For an operation $g : D^m \to D$, let $g^\pi : D^m \to D$ be the following operation:

$$g^\pi(\alpha) = g(\alpha^\pi) \tag{6}$$

- For a mapping $\mathbf{g} : D^m \to D^m$, let $\mathbf{g}^s : D^m \to D^m$ be the following mapping:

$$\mathbf{g}^s(\alpha) = [\mathbf{g}(\alpha)]^s \tag{7}$$

The last definition can also be expressed as

$$\mathbf{g}^s = (s \circ \mathbf{g}_{-1}, \ldots, s \circ \mathbf{g}_{-m}) \tag{8}$$

where $\mathbf{g}_{-i} : D^m \to D^{m-1}$ is the sequence of $m-1$ operations obtained from $\mathbf{g} = (g_1, \ldots, g_m)$ by removing the $i$-th operation.

Let $\mathbb{1}$ be the identity mapping $D^m \to D^m$, and let $V = \{\mathbb{1}^{s_1 \ldots s_k} \mid s_1, \ldots, s_k \in supp(\omega), k \geq 0\}$ be the set of all mappings that can be generated from $\mathbb{1}$.

**Proposition 8.** *Every $\mathbf{g} = (g_1, \ldots, g_m) \in V$ satisfies the following:*

$$(g_1^\pi, \ldots, g_m^\pi) = (g_{\pi(1)}, \ldots, g_{\pi(m)}) \qquad \forall \text{ permutation } \pi \tag{9}$$

Thus, permuting the arguments of $g_i(\cdot, \ldots, \cdot)$ gives a mapping which is also present in the sequence $\mathbf{g}$, possibly at a different position.

*Proof.* Checking that $\mathbb{1}$ satisfies (9) is straightforward. Let us prove that for any $\mathbf{g} : D^m \to D^m$ satisfying (9) and for any symmetric operation $s \in \mathcal{O}^{m-1}$ mapping $\mathbf{g}^s$ also satisfies (9). Consider $i \in [1, m]$. We need to show that $(s \circ \mathbf{g}_{-i})^\pi = s \circ \mathbf{g}_{-\pi(i)}$. For each $\alpha \in D^m$ we have

$$\begin{aligned}
(s \circ \mathbf{g}_{-i})^\pi(\alpha) = s \circ \mathbf{g}_{-i}(\alpha^\pi) &= s(g_1(\alpha^\pi), \ldots, g_{i-1}(\alpha^\pi), g_{i+1}(\alpha^\pi), \ldots, g_m(\alpha^\pi)) \\
&= s(g_1^\pi(\alpha), \ldots, g_{i-1}^\pi(\alpha), g_{i+1}^\pi(\alpha), \ldots, g_m^\pi(\alpha)) \\
&= s(g_{\pi(1)}(\alpha), \ldots, g_{\pi(i-1)}(\alpha), g_{\pi(i+1)}(\alpha), \ldots, g_{\pi(m)}(\alpha)) = s \circ \mathbf{g}_{-\pi(i)}(\alpha)
\end{aligned}$$

$\square$

**Graph on mappings** Let us define a directed weighted graph $G = (V, E, w)$ with the set of edges $E = \{(\mathbf{g}, \mathbf{g}^s) \mid \mathbf{g} \in V, s \in supp(\omega)\}$ and positive weights $w(\mathbf{g}, \mathbf{h}) = \sum_{s \in supp(\omega), \mathbf{h} = \mathbf{g}^s} \omega(s)$ for $(\mathbf{g}, \mathbf{h}) \in E$. Clearly, we have

$$\sum_{(\mathbf{g},\mathbf{h}) \in E} w(\mathbf{g}, \mathbf{h}) = 1 \qquad \mathbf{g} \in V \tag{10}$$

We define $\mathbb{H}[G]$ to be the set of strongly connected components $H \subseteq V$ of $G$ which are sinks, i.e. all edges in $G$ from $H$ lead to vertices in $H$. We also denote $\widehat{H} = \bigcup_{H \in \mathbb{H}[G]} H \subseteq V$.

## 3.1 Proof overview: main theorems

From now on we fix a function $f \in \Gamma$ of arity $n$. Recall that we defined $f^m(x^1, \ldots, x^m) = \frac{1}{m} \sum_{i \in [1,m]} f(x^i)$. For a mapping $\mathbf{g} \in \mathcal{O}^{m \to m}$ we define

$$Range_n(\mathbf{g}) = \{\mathbf{g}(x^1, \ldots, x^m) \mid x^1, \ldots, x^m \in D^n\}$$

We will prove the following.

**Theorem 9.** *There exists a fractional polymorphism $\rho$ of $\Gamma$ of arity $m \to m$ with $supp(\rho) \subseteq \widehat{H}$.*

**Theorem 10.** *Let $\hat{\mathbf{g}}$ be a mapping in $\widehat{H}$ and $\mathbf{p} \in \mathcal{O}^{m \to m}$ be any mapping such that $\mathbf{p}(\alpha)$ is a permutation of $\alpha$ for all $\alpha \in D^m$. For any $(x^1, \ldots, x^m) \in Range_n(\hat{\mathbf{g}})$ there holds $f^m(x^1, \ldots, x^m) = f^m(\mathbf{p}(x^1, \ldots, x^m))$.*

This will imply Theorem 7. Indeed, we can construct an $m$-ary symmetric fractional polymorphism of $\Gamma$ as follows. Take vector $\rho$ from theorem 9, take a symmetric mapping $\mathbf{p} \in \mathcal{O}^{m \to m}$ satisfying the condition of theorem 10, and define a fractional polymorphism of arity $m \to m$

$$\rho' = \sum_{\mathbf{g} \in supp(\rho)} \rho(\mathbf{g}) \chi_{\mathbf{p} \circ \mathbf{g}}$$

Function $f$ admits $\rho'$ since for any labelings $x^1, \ldots, x^m \in D^n$ there holds

$$\sum_{\mathbf{h} \in supp(\rho')} \rho'(\mathbf{h}) f^m(\mathbf{h}(x^1, \ldots, x^m)) = \sum_{\mathbf{g} \in supp(\rho)} \rho(\mathbf{g}) f^m(\mathbf{p}(\mathbf{g}(x^1, \ldots, x^m)))$$

$$= \sum_{\mathbf{g} \in supp(\rho)} \rho(\mathbf{g}) f^m(\mathbf{g}(x^1, \ldots, x^m)) \leq f^m(x^1, \ldots, x^m)$$

Note, for any $\mathbf{h} = (h_1, \ldots, h_m) \in supp(\rho')$ operations $h_1, \ldots, h_m$ are symmetric. Indeed, we have $\mathbf{h} = \mathbf{p} \circ \mathbf{g}$ for some $\mathbf{g} \in V$. If $\alpha \in D^m$ and $\pi$ is a permutation of size $m$ then $\mathbf{h}(\alpha^\pi) = \mathbf{p}(\mathbf{g}(\alpha^\pi)) = \mathbf{p}([\mathbf{g}(\alpha)]^\pi) = \mathbf{p}(\mathbf{g}(\alpha)) = \mathbf{h}(\alpha)$ which implies the claim.

We can finally apply transformation (3) to vector $\rho'$ to get a symmetric fractional polymorphism of arity $m$.

A proof of Theorem 9 is given in section 3.2. To prove Theorem 10, we will need an auxiliary result. Let us fix a connected component $H \in \mathbb{H}[G]$, and denote $I = H \times [1, m]$. Given labelings $x^1, \ldots, x^m$, we define labelings $x^{\mathbf{g}i}$ for all $(\mathbf{g}, i) \in I$ via

$$(x^{\mathbf{g}1}, \ldots, x^{\mathbf{g}m}) = \mathbf{g}(x^1, \ldots, x^m) \tag{11}$$

Note that $x^{\mathbf{g}i}$ is a function of $(x^1, \ldots, x^m)$; for brevity of notation, this dependence is not shown. For a vector $\lambda \in \mathbb{R}^H$ and an index $i \in [1, m]$ we define function $F_i^\lambda$ via

$$F_i^\lambda(x^1, \ldots, x^m) = \sum_{\mathbf{g} \in H} \lambda_{\mathbf{g}} f(x^{\mathbf{g}i}) \qquad \forall x^1, \ldots, x^m \in D^n \tag{12}$$

**Theorem 11.** *Consider $H \in \mathbb{H}[G]$ and $x^1, \ldots, x^m \in D^n$.*
*(a) There holds $f^m(x^{\mathbf{g}'1}, \ldots, x^{\mathbf{g}'m}) = f^m(x^{\mathbf{g}''1}, \ldots, x^{\mathbf{g}''m})$ for all $\mathbf{g}', \mathbf{g}'' \in H$.*
*(b) There exists a probability distribution $\lambda$ over $H$ (that depends only on $H$ and $\omega$) such that $F_{i'}^\lambda(x^1, \ldots, x^m) = F_{i''}^\lambda(x^1, \ldots, x^m)$ for all $i', i'' \in [1, m]$.*

We prove this theorem in section 3.3; using this result, we then prove Theorem 10 in section 3.4.

## 3.2 Proof of theorem 9

We will rely on the following fact.

**Proposition 12.** *Suppose that $\rho$ is a fractional polymorphism of $\Gamma$ of arity $m \to m$ and $\mathbf{g} \in supp(\rho)$ with $\lambda = \rho(\mathbf{g})$. Then vector*

$$\rho' = \rho - \lambda \cdot \chi_{\mathbf{g}} + \lambda \sum_{s \in supp(\omega)} \omega(s) \chi_{\mathbf{g}^s} \tag{13}$$

*is also a fractional polymorphism of $\Gamma$ of arity $m \to m$.*

*Proof.* Denote the sum on the RHS of (13) as $\rho_{\mathbf{g}}$, so that $\rho' = \rho - \lambda \cdot \chi_{\mathbf{g}} + \lambda \cdot \rho_{\mathbf{g}}$. Consider function $f \in \Gamma$ of arity $n$ and labelings $x^1, \ldots, x^m \in D^n$, and denote $(y^1, \ldots, y^m) = \mathbf{g}(x^1, \ldots, x^m)$. We can write

$$\begin{aligned}
\sum_{\mathbf{h} \in supp(\rho_{\mathbf{g}})} \rho_{\mathbf{g}}(\mathbf{h}) f^m(\mathbf{h}(x^1, \ldots, x^m)) &= \sum_{s \in supp(\omega)} \omega(s) f^m(\mathbf{g}^s(x^1, \ldots, x^m)) \\
&= \sum_{s \in supp(\omega)} \omega(s) \frac{1}{m} \sum_{i \in [1,m]} f((y^1, \ldots, y^m)_{-i}) \\
&= \frac{1}{m} \sum_{i \in [1,m]} \sum_{s \in supp(\omega)} \omega(s) f((y^1, \ldots, y^m)_{-i}) \\
&\leq \frac{1}{m} \sum_{i \in [1,m]} f^{m-1}((y^1, \ldots, y^m)_{-i})) \\
&= f^m(y^1, \ldots, y^m) = f^m(\mathbf{g}(x^1, \ldots, x^m))
\end{aligned}$$

Therefore,

$$\begin{aligned}
\sum_{\mathbf{h} \in supp(\rho')} \rho'(\mathbf{h}) f^m(\mathbf{h}(x^1, \ldots, x^m)) &= \sum_{\mathbf{h} \in supp(\rho) - \{\mathbf{g}\}} \rho(\mathbf{h}) f^m(\mathbf{h}(x^1, \ldots, x^m)) + \lambda \sum_{\mathbf{h} \in supp(\rho_{\mathbf{g}})} \rho_{\mathbf{g}}(\mathbf{h}) f^m(\mathbf{h}(x^1, \ldots, x^m)) \\
&\leq \sum_{\mathbf{h} \in supp(\rho) - \{\mathbf{g}\}} \rho(\mathbf{h}) f^m(\mathbf{h}(x^1, \ldots, x^m)) + \rho(\mathbf{g}) f^m(\mathbf{g}(x^1, \ldots, x^m)) \\
&= \sum_{\mathbf{h} \in supp(\rho)\}} \rho(\mathbf{h}) f^m(\mathbf{h}(x^1, \ldots, x^m)) \leq f^m(x^1, \ldots, x^m)
\end{aligned}$$

□

We can now prove Theorem 9. Let $\Omega$ be the set of fractional polymorphisms $\rho$ of $\Gamma$ of arity $m \to m$ with $supp(\rho) \subseteq V$; it is non-empty since $\chi_{\mathbb{1}} \in \Omega$. Let us pick $\rho \in \Omega$ with the maximum value of $\rho(\widehat{H}) \doteq \sum_{\mathbf{g} \in \widehat{H}} \rho(\mathbf{g})$. (Clearly, $\Omega$ is a compact set, so the maximum is attained by some vector in $\Omega$). We claim that $supp(\rho) \in \widehat{H}$. Indeed, suppose that there exists $\mathbf{g} \in supp(\rho)$ with $\mathbf{g} \notin \widehat{H}$. By the definition of $\widehat{H}$ there exists a path in $G$ from $\mathbf{g}$ to a vertex $\mathbf{h} \in \widehat{H}$, i.e. $\mathbf{h} = \mathbf{g}^{s_1 \ldots s_k}$ for some $s_1, \ldots, s_k \in supp(\omega)$. Therefore, we can repeatedly modify $\rho$ by applying a sequence of transformations (13) to the current vector $\rho$ and some $\mathbf{g}_i \in supp(\rho) - \widehat{H}$ to get vector $\rho'$ with larger weight $\rho'(\widehat{H})$, using at most $k$ steps. This contradicts to the choice of $\rho$. Theorem 9 is proved.

We mention that a similar argument can be used as an alternative to the "tree cutting" argument in the proof of Theorem 4.4 in [20] (and vice versa). A yet another proof technique was used in [21].

## 3.3 Proof of Theorem 11

First, we make the following observation.

**Proposition 13.** *If* $\mathbf{h} = \mathbf{g}^s$ *where* $\mathbf{g} \in H$, $s \in supp(\omega)$ *then* $x^{\mathbf{h}i} = s((x^{\mathbf{g}1},\ldots,x^{\mathbf{g}m})_{-i})$ *for* $i \in [1,m]$ *where* $(x^{\mathbf{g}1},\ldots,x^{\mathbf{g}m})_{-i}$ *is the sequence of* $m-1$ *labelings obtained by removing the* $i$-*th labeling.*

*Proof.* Consider node $v \in [1,n]$, and denote $\alpha = (x_v^1, \ldots, x_v^m)$, $\beta = (x_v^{\mathbf{g}1}, \ldots, x_v^{\mathbf{g}m})$, $\gamma = (x_v^{\mathbf{h}1}, \ldots, x_v^{\mathbf{h}m})$. By definition (11), $\beta = \mathbf{g}(\alpha)$ and $\gamma = \mathbf{h}(\alpha)$. Therefore, $\gamma = \mathbf{g}^s(\alpha) = [\mathbf{g}(\alpha)]^s = \beta^s$. In other words, the $i$-th component of $\gamma$ equals $s(\beta_{-i})$, which is what we needed to show. □

We will show that for fixed distinct mappings $\mathbf{g}', \mathbf{g}'' \in H$ there holds

$$\sum_{i \in [1,m]} f(x^{\mathbf{g}'i}) - \sum_{i \in [1,m]} f(x^{\mathbf{g}''i}) \leq 0 \tag{14a}$$

and that there exists a probability distribution $\lambda$ over $H$ such that for fixed distinct indices $i', i'' \in [1,m]$ there holds

$$\sum_{\mathbf{g} \in H} \lambda_{\mathbf{g}} f(x^{\mathbf{g}i'}) - \sum_{\mathbf{g} \in H} \lambda_{\mathbf{g}} f(x^{\mathbf{g}i''}) \leq 0 \tag{14b}$$

Clearly, this will imply Theorem 11.

To prove these facts, we will use the following strategy. For each $(\mathbf{g}, i) \in I$ let us write the polymorphism inequality for labelings $(x^{\mathbf{g}1},\ldots,x^{\mathbf{g}m})_{-i}$:

$$\sum_{s \in supp(\omega)} \omega(s) f(s((x^{\mathbf{g}1},\ldots,x^{\mathbf{g}m})_{-i})) \leq \frac{1}{m-1} \sum_{j \in [1,m]-\{i\}} f(x^{\mathbf{g}j})$$

Let us multiply this inequality by weight $\lambda_{\mathbf{g}i} \geq 0$ (to be defined later), and apply Proposition 13:

$$\lambda_{\mathbf{g}i} \sum_{s \in supp(\omega), \mathbf{h}=\mathbf{g}^s} \omega(s) f(x^{\mathbf{h}i}) - \frac{\lambda_{\mathbf{g}i}}{m-1} \sum_{j \in [1,m]-\{i\}} f(x^{\mathbf{g}j}) \leq 0 \qquad \forall (\mathbf{g}, i) \in I$$

Summing these inequalities over $(\mathbf{g}, i) \in I$ gives

$$\sum_{(\mathbf{g},i) \in I} \left[ \sum_{s \in supp(\omega), \mathbf{g}=\mathbf{h}^s} \omega(s) \lambda_{\mathbf{h}i} - \sum_{j \in [1,m]-\{i\}} \frac{\lambda_{\mathbf{g}j}}{m-1} \right] f(x^{\mathbf{g}i}) \leq 0 \tag{15}$$

Parts (a,b) of Lemma 14 together with Remark 1 below show that coefficients $\lambda_{\mathbf{g}i}$ can be chosen in such a way that the last inequality becomes equivalent to (14a) and (14b) respectively, thus proving Theorem 11.

**Lemma 14.** *(a) There exists vector* $\lambda \in \mathbb{R}_{\geq 0}^I$ *that satisfies*

$$\sum_{(\mathbf{h},\mathbf{g}) \in E} w(\mathbf{h},\mathbf{g}) \lambda_{\mathbf{h}i} - \sum_{j \in [1,m]-\{i\}} \frac{\lambda_{\mathbf{g}j}}{m-1} = c_{\mathbf{g}} \qquad \forall (\mathbf{g}, i) \in I \tag{16}$$

*where* $c_{\mathbf{g}} = [\mathbf{g} = \mathbf{g}'] - [\mathbf{g} = \mathbf{g}'']$ *and* $[\cdot]$ *is the Iverson bracket: it equals* 1 *if the argument is true, and* 0 *otherwise.*
*(b) There exists vector* $\lambda \in \mathbb{R}_{\geq 0}^{I \cup H}$ *that satisfies*

$$\sum_{(\mathbf{h},\mathbf{g}) \in E} w(\mathbf{h},\mathbf{g}) \lambda_{\mathbf{h}i} - \sum_{j \in [1,m]-\{i\}} \frac{\lambda_{\mathbf{g}j}}{m-1} = c_i \lambda_{\mathbf{g}} \qquad \forall (\mathbf{g}, i) \in I \tag{17a}$$

$$\sum_{\mathbf{g} \in H} \lambda_{\mathbf{g}} = 1 \tag{17b}$$

*where* $c_i = [i = i'] - [i = i'']$.

**Remark 1** Note that vector $\lambda$ in part (b) depends on the pair $(i', i'')$; let us denote it as $\lambda^{i'i''}$. To establish Theorem 11(b), it suffices to show that vectors $\lambda^{i'i''}$ in Lemma 14(b) can be chosen in such a way that for a given $\mathbf{g} \in H$ components $\lambda^{i'i''}_{\mathbf{g}}$ are the same for all pairs $(i', i'')$. This can be done as follows. Take vector $\lambda^{12}$ constructed in Lemma 14(b). For a pair of distinct indices $(i', i'') \neq (1, 2)$ select permutation $\pi$ of $[1, m]$ with $\pi(i') = 1$, $\pi(i'') = 2$, and define vector $\lambda^{i'i''}$ via

$$\lambda^{i'i''}_{\mathbf{g}} = \lambda^{12}_{\mathbf{g}} \qquad \forall \mathbf{g} \in H \qquad\qquad \lambda^{i'i''}_{\mathbf{g}i} = \lambda^{12}_{\mathbf{g}\pi(i)} \qquad \forall (\mathbf{g}, i) \in I$$

Clearly, vector $\lambda^{i'i''}$ satisfies conditions of Lemma 14(b) for the pair $(i', i'')$. Thus, Lemma 14 indeed implies Theorem 11. The proof of the lemma is given below.

*Proof.* **Part (a)** Suppose the claim does not hold. By Farkas lemma there exists vector $y \in \mathbb{R}^I$ such that

$$\sum_{(\mathbf{g},i) \in I} c_{\mathbf{g}} y_{\mathbf{g}i} \;<\; 0 \tag{18a}$$

$$\sum_{(\mathbf{g},\mathbf{h}) \in E} w(\mathbf{g},\mathbf{h}) y_{\mathbf{h}i} - \sum_{j \in [1,m]-\{i\}} \frac{y_{\mathbf{g}j}}{m-1} \;\geq\; 0 \qquad \forall (\mathbf{g}, i) \in I \tag{18b}$$

Denote $u_{\mathbf{g}} = \sum_{i \in [1,m]} y_{\mathbf{g}i}$. Summing inequalities (18b) over $i \in [1, m]$ gives

$$\sum_{(\mathbf{g},\mathbf{h}) \in E} w(\mathbf{g},\mathbf{h}) u_{\mathbf{h}} - u_{\mathbf{g}} \;\geq\; 0 \qquad \forall \mathbf{g} \in H \tag{19}$$

Denote $H^* = \arg\max\{u_{\mathbf{g}} \mid \mathbf{g} \in H\}$. From (10) and (19) we conclude that $\mathbf{g} \in H^*$ implies $\mathbf{h} \in H^*$ for all $(\mathbf{g}, \mathbf{h}) \in E$. Therefore, $H^* = H$ (since $H$ is a strongly connected component of $G$).

We showed that $u_{\mathbf{g}} = C$ for all $\mathbf{g} \in H$ where $C \in \mathbb{R}$ is some constant. But then the expression on the LHS of (18a) equals $C - C = 0$ - a contradiction.

**Part (b)** Suppose the claim does not hold. By Farkas lemma there exist vector $y \in \mathbb{R}^I$ and scalar $z \in \mathbb{R}$ such that

$$z \;<\; 0 \tag{20a}$$

$$z - \sum_{i \in [1,m]} c_i y_{\mathbf{g}i} \;\geq\; 0 \qquad \forall \mathbf{g} \in H \tag{20b}$$

$$\sum_{(\mathbf{g},\mathbf{h}) \in E} w(\mathbf{g},\mathbf{h}) y_{\mathbf{h}i} - \sum_{j \in [1,m]-\{i\}} \frac{y_{\mathbf{g}j}}{m-1} \;\geq\; 0 \qquad \forall (\mathbf{g}, i) \in I \tag{20c}$$

Denote $u_{\mathbf{g}} = \sum_{i \in [1,m]} y_{\mathbf{g}i}$. Using the same argument as in part (a) we conclude that $u_{\mathbf{g}} = C$ for all $\mathbf{g} \in H$ where $C \in \mathbb{R}$ is some constant. We can assume w.l.o.g. that this constant is zero. Indeed, this can be achieved by subtracting $C/m$ from values $y_{\mathbf{g}i}$ for all $(\mathbf{g}, i) \in I$ with $\mathbf{g} \in H$; it can be checked (using eq. (10)) that this operation preserves inequalities (20). We thus have

$$\sum_{j \in [1,m]-\{i\}} y_{\mathbf{g}j} \;=\; -y_{\mathbf{g}i} \qquad \forall (\mathbf{g}, i) \in I \tag{21}$$

Substituting this into (20c) gives

$$\sum_{(\mathbf{g},\mathbf{h}) \in E} w(\mathbf{g},\mathbf{h}) y_{\mathbf{h}i} + \frac{y_{\mathbf{g}i}}{m-1} \;\geq\; 0 \qquad \forall (\mathbf{g}, i) \in I \tag{22}$$

Consider $i' \in [1, m]$. Summing (22) over $i \in [1, m] - \{i'\}$ and then using (21) yields

$$\sum_{(\mathbf{g},\mathbf{h}) \in E} w(\mathbf{g},\mathbf{h})(-y_{\mathbf{h}i'}) + \frac{-y_{\mathbf{g}i'}}{m-1} \;\geq\; 0 \qquad \forall (\mathbf{g}, i') \in I \tag{23}$$

Combining (22) and (23) gives

$$\sum_{(\mathbf{g},\mathbf{h})\in E} w(\mathbf{g},\mathbf{h})y_{\mathbf{h}i} + \frac{y_{\mathbf{g}i}}{m-1} = 0 \quad \forall (\mathbf{g},i) \in I \quad (24)$$

Denote $r_{\mathbf{g}} = \sum_{i\in[1,m]} c_i y_{\mathbf{g}i}$ for $\mathbf{g} \in H$. Summing (24) over $i \in [1,m]$ with appropriate coefficients gives

$$\sum_{(\mathbf{g},\mathbf{h})\in E} w(\mathbf{g},\mathbf{h})r_{\mathbf{h}} + \frac{r_{\mathbf{g}}}{m-1} = 0 \quad \forall \mathbf{g} \in H \quad (25)$$

From (20a) and (20b) we conclude that $r_{\mathbf{g}} < 0$ for all $\mathbf{g} \in H$, and thus eq. (25) cannot hold - a contradiction. $\square$

### 3.4 Proof of Theorem 10

Let $H \in \mathbb{H}[G]$ be the strongly connected component that contains $\hat{\mathbf{g}}$, and let $\lambda \in \mathbb{R}^H_{\geq 0}$ be a vector constructed in Theorem 11(b). We denote $F^\lambda(x^1,\ldots,x^m) = F_i^\lambda(x^1,\ldots,x^m)$ for $i \in [1,m]$.

**Lemma 15.** *The following transformation does not change $F^\lambda(x^1,\ldots,x^m)$: pick node $v \in [1,m]$ and permute values $(x_v^1,\ldots,x_v^m)$.*

*Proof.* It suffices to prove the claim for a permutation which swaps values $x_v^i$ and $x_v^j$ for $i,j \in [1,m]$ (since any other permutation can be obtained by repeatedly applying such swaps). Since $m \geq 3$ there exists index $k \in [1,m] - \{i,j\}$. Using Proposition 8, it can be checked that the swap above does not affect labelings $x^{\mathbf{g}k}$ in (11) for $\mathbf{g} \in H$, and therefore $F_k^\lambda(x^1,\ldots,x^m)$ does not change. $\square$

**Lemma 16.** *If $(x^1,\ldots,x^m) \in Range_n(\hat{\mathbf{g}})$ then $(x^1,\ldots,x^m) = (x^{\mathbf{g}1},\ldots,x^{\mathbf{g}m})$ for some $\mathbf{g} \in H$.*

*Proof.* We need to show that there exists $\mathbf{g} \in H$ with $\mathbf{g} \circ \hat{\mathbf{g}} = \hat{\mathbf{g}}$.

Note that $\mathbb{1}^{s_1 \ldots s_k} \circ \mathbf{h} = \mathbf{h}^{s_1 \ldots s_k}$ for any $s_1,\ldots,s_k \in supp(\omega)$ and $\mathbf{h} \in \mathcal{O}^{m \to m}$, since

$$[\mathbb{1}^{s_1 \ldots s_k} \circ \mathbf{h}](\alpha) = \mathbb{1}^{s_1 \ldots s_k}(\mathbf{h}(\alpha)) = [\mathbb{1}(\mathbf{h}(\alpha))]^{s_1 \ldots s_k} = [\mathbf{h}(\alpha)]^{s_1 \ldots s_k} = \mathbf{h}^{s_1 \ldots s_k}(\alpha) \quad \forall \alpha \in D^m$$

Therefore, conditions $\mathbf{g} \in V$, $\mathbf{h} \in H$ imply that $\mathbf{g} \circ \mathbf{h} \in H$ (since $\mathbf{g}$ can be written as $\mathbf{g} = \mathbb{1}^{s_1 \ldots s_k}$ and there are no edges leaving $H$).

Since $H$ is strongly connected, there is a path in $G$ from $\hat{\mathbf{g}} \circ \hat{\mathbf{g}} \in H$ to $\hat{\mathbf{g}} \in H$, i.e. $[\hat{\mathbf{g}} \circ \hat{\mathbf{g}}]^{s_1 \ldots s_k} = \hat{\mathbf{g}}$ for some $s_1,\ldots,s_k \in supp(\omega)$. Equivalently, $\mathbf{h} \circ \hat{\mathbf{g}} \circ \hat{\mathbf{g}} = \hat{\mathbf{g}}$ where $\mathbf{h} = \mathbb{1}^{s_1 \ldots s_k}$. It can be checked that mapping $\mathbf{g} = \mathbf{h} \circ \hat{\mathbf{g}}$ has the desired properties. $\square$

**Lemma 17.** *If $(x^1,\ldots,x^m) \in Range_n(\hat{\mathbf{g}})$ then $f^m(x^1,\ldots,x^m) = F^\lambda(x^1,\ldots,x^m)$.*

*Proof.* From Theorem 11(a) and Lemma 16 we get that $f^m(x^{\mathbf{g}1},\ldots,x^{\mathbf{g}m}) = f^m(x^1,\ldots,x^m)$ for all $\mathbf{g} \in H$. Using this fact and the definition of $F_i^\lambda(\cdot)$, we can write

$$F^\lambda(x^1,\ldots,x^m) = \frac{1}{m}\sum_{i\in[1,m]} F_i^\lambda(x^1,\ldots,x^m) = \frac{1}{m}\sum_{\mathbf{g}\in H} \lambda_{\mathbf{g}} \sum_{i\in[1,m]} f(x^{\mathbf{g}i})$$

$$= \sum_{\mathbf{g}\in H} \lambda_{\mathbf{g}} f^m(x^{\mathbf{g}1},\ldots,x^{\mathbf{g}m}) = \sum_{\mathbf{g}\in H} \lambda_{\mathbf{g}} f^m(x^1,\ldots,x^m) = f^m(x^1,\ldots,x^m)$$

$\square$

To establish Theorem 10, it remains to prove that condition $(x^1,\ldots,x^m) \in Range_n(\hat{\mathbf{g}})$ implies $\mathbf{p}(x^1,\ldots,x^m) \in Range_n(\hat{\mathbf{g}})$. This proof follows mechanically from Proposition 8, and is omitted.

## 4 STP multimorphisms

In this section we consider Symmetric Tournament Pairs (STP) multimorphisms [3].

**Definition 18.** *(a) A pair of operations $\langle \sqcap, \sqcup \rangle$ with $\sqcap, \sqcup : D \times D \to D$ is called an STP multimorphism if*

$$a \sqcap b = b \sqcap a, \quad a \sqcup b = b \sqcup a \qquad \forall a, b \in D \qquad (commutativity) \qquad (26a)$$
$$\{a \sqcap b, a \sqcup b\} = \{a, b\} \qquad \forall a, b \in D \qquad (conservativity) \qquad (26b)$$

*(b) Pair $\langle \sqcap, \sqcup \rangle$ is called a* submodularity multimorphism *if there exists a total order on $D$ for which $a \sqcap b = \min\{a, b\}$, $a \sqcup b = \max\{a, b\}$ for all $a, b \in D$.*
*(c) Language $\Gamma$ admits $\langle \sqcap, \sqcup \rangle$ if every function $f \in \Gamma$ of arity $n$ satisfies*

$$f(x \sqcap y) + f(x \sqcup y) \leq f(x) + f(y) \qquad \forall x, y \in D^n \qquad (27)$$

It has been shown in [3] that if $\Gamma$ admits an STP multimorphism then $VCSP(\Gamma)$ can be solved in polynomial time. STP multimorphisms also appeared in the dichotomy result of [12]:

**Theorem 19.** *Suppose a finite-valued language $\Gamma$ is* conservative, *i.e. it contains all possible unary cost functions $u : D \to \{0, 1\}$. Then $\Gamma$ either admits an STP multimorphism or it is NP-hard.*

To goal of this section is to prove the following.

**Theorem 20.** *If a finite-valued language $\Gamma$ admits an STP multimorphism then it also admits a submodularity multimorphism.*

This fact is already known; in particular, footnote 2 in [12] mentions that this result is implicitly contained in [3], and sketches a proof strategy. However, to our knowledge a formal proof has never appeared in the literature. We now fill this gap. Our proof is different from the one suggested in [12], and inspired some of the proof techniques used in the main part of this paper.

### 4.1 Proof of Theorem 20

Consider a directed graph $G = (D, E)$. We say that $G$ is *complete* if for each pair of distinct labels $a, b \in D$ exactly one of the edges $(a, b), (b, a)$ belong to $E$. We define a one-to-one correspondence between STP multimorphisms $\langle \sqcap, \sqcup \rangle$ and complete graphs $G = (D, E)$ as follows:

$$(a, b) \in E \quad \Leftrightarrow \quad (a \sqcap b, a \sqcup b) = (a, b) \qquad \forall a, b \in D, a \neq b$$

It can be seen that $\langle \sqcap, \sqcup \rangle$ is a submodularity multimorphism iff the corresponding graph $G$ is acyclic.

**Lemma 21.** *Suppose a finite-valued language $\Gamma$ admits an STP multimorphism $\langle \sqcap, \sqcup \rangle$ corresponding to a directed graph $G = (D, E)$, and suppose that $G$ has a 3-cycle: $(a, b), (b, c), (c, a) \in E$. Let $\hat{G}$ be the graph obtained from $G$ by reversing the orientation of edge $(a, b)$, and let $\langle \hat{\sqcap}, \hat{\sqcup} \rangle$ be the corresponding STP multimorphism. Then $\Gamma$ admits $\langle \hat{\sqcap}, \hat{\sqcup} \rangle$.*

*Proof.* Let $\langle \wedge, \vee \rangle$ be the following multimorphism:

$$(x \wedge y, x \vee y) = \begin{cases} (x, y) & \text{if } (x, y) \in \{(a, b), (b, a)\} \\ (x \sqcap y, x \sqcup y) & \text{if } (x, y) \notin \{(a, b), (b, a)\} \end{cases}$$

First, we will prove that $\Gamma$ admits $\langle \wedge, \vee \rangle$ (step 1), and then prove that $\Gamma$ admits $\langle \hat{\sqcap}, \hat{\sqcup} \rangle$ (step 2). We fix below function $f \in \Gamma$ of arity $n$ and labelings $x, y \in D^n$.

**Step 1** Let us define labelings $x', y' \in D^n$ via

$$(x'_v, y'_v) = \begin{cases} (x_v, x_v \sqcap y_v) & \text{if } (x_v, y_v) \neq (b, a) \\ (c, c) & \text{if } (x_v, y_v) = (b, a) \end{cases} \quad \forall v \in [1, n]$$

It can be checked that the following identities hold:

$$x' \sqcap y = y' \qquad x \sqcup y' = x' \tag{28a}$$
$$x \sqcap y' = x \wedge y \qquad x' \sqcup y = x \vee y \tag{28b}$$

Let us write multimorphism inequalities for pairs $(x', y)$ and $(x, y')$:

$$\underline{f(x' \sqcap y)} + f(x' \sqcup y) \leq \underline{f(x')} + f(y) \tag{29a}$$
$$f(x \sqcap y') + \underline{f(x \sqcup y')} \leq f(x) + \underline{f(y')} \tag{29b}$$

Summing (29a) and (29b), cancelling terms using (28a), and then substituting expressions using (28b) gives

$$f(x \wedge y) + f(x \vee y) \leq f(x) + f(y) \tag{30}$$

**Step 2** Let us define labelings $x', y' \in D^n$ via

$$(x'_v, y'_v) = \begin{cases} (x_v \wedge y_v, y_v) & \text{if } (x_v, y_v) \neq (a, b) \\ (c, c) & \text{if } (x_v, y_v) = (a, b) \end{cases} \quad \forall v \in [1, n]$$

It can be checked that the following identities hold:

$$x' \vee y = y' \qquad x \wedge y' = x' \tag{31a}$$
$$x \vee y' = x \,\hat{\sqcup}\, y \qquad x' \wedge y = x \,\hat{\sqcap}\, y \tag{31b}$$

Let us write multimorphism inequalities for pairs $(x', y)$ and $(x, y')$:

$$f(x' \wedge y) + \underline{f(x' \vee y)} \leq \underline{f(x')} + f(y) \tag{32a}$$
$$\underline{f(x \wedge y')} + f(x \vee y') \leq f(x) + \underline{f(y')} \tag{32b}$$

Summing (32a) and (32b), cancelling terms using (31a), and then substituting expressions using (31b) gives

$$f(x \,\hat{\sqcap}\, y) + f(x \,\hat{\sqcup}\, y) \leq f(x) + f(y) \tag{33}$$

□

We call the operation of reversing the orientation of edge $(a, b) \in E$ in a graph $G = (D, E)$ a *valid flip* if $(a, b)$ belongs to a 3-cycle. To prove Theorem 20, it thus suffices to show the following:

- For any complete graph $G$ there exists a sequence of valid flips that makes it acyclic.

Such sequence can be constructed as follows: (1) start with a subset $B \subseteq D$ with $|B| = 2$; (2) perform valid flips in $G[B]$ to make it acyclic, where $G[B] = (B, E[B])$ is the subgraph of $G$ induced by $B$; (3) if $B \neq D$, add a node $c \in D - B$ to $B$ and repeat step 2. The lemma below shows how to implement step 2.

**Lemma 22.** *Suppose that $G = (D, E)$ is a complete graph, $D = B \cup \{c\}$ with $c \notin B$ and subgraph $G[B]$ is acyclic. Then there exists a sequence of valid flips that makes $G$ acyclic.*

*Proof.* Suppose that $G$ has a cycle $\mathcal{C}$, then it must pass through $c$ (since $G[B]$ is acyclic): $\mathcal{C} = \ldots \to b \to c \to a \to \ldots$. Since there is a path from $a$ to $b$ in $G[B]$, we must have $(a, b) \in E$ (again, due to acyclicity of $G[B]$). Thus, $c \to a \to b \to c$ is a 3-cycle in $G$.

Let us repeat the following procedure while possible: pick such cycle and flip edge $(c, a)$ to $(a, c)$. This operation decreases the number of edges in $G$ coming out of $c$. Therefore, it must terminate after a finite number of steps and yield an acyclic graph $G$. □

## Acknowledgements

I thank Andrei Krokhin for helpful discussions and for communicating the result of Raghavendra [16] about cyclic fractional polymorphisms.